\documentclass[12pt]{JHEP3}

\usepackage{amsmath}

\usepackage{amsfonts}

\usepackage{mathrsfs}

\usepackage{cite}

\newcommand{\beqs}{\begin{equation*}}

\newcommand{\beq}{\begin{equation}}

\newcommand{\eeqs}{\end{equation*}}

\newcommand{\eeq}{\end{equation}}

\newcommand{\beqas}{\begin{eqnarray*}}

\newcommand{\beqa}{\begin{eqnarray}}

\newcommand{\eeqas}{\end{eqnarray*}}

\newcommand{\eeqa}{\end{eqnarray}}

\newcommand{\eps}{\varepsilon}

\newcommand{\ga}{\gamma}

\newcommand{\de}{\delta}

\newcommand{\Ga}{\Gamma}

\newcommand{\blist}{\begin{itemize}}

\newcommand{\elist}{\end{itemize}}

\providecommand{\href}[2]{#2}

\newcommand\MM{\mathcal{M}}

\newcommand\dM{\partial \MM}

\title{Erratum to `Instability in cosmological topologically massive gravity at the chiral point', \href{http://arxiv.org/abs/0805.2610}{{\tt 0805.2610}}}

\author{Sabine Ertl, Daniel Grumiller and Niklas Johansson\\
         Institute for Theoretical Physics,
         Vienna University of Technology\\
         Wiedner Hauptstr.~8-10/136, A-1040 Vienna, Austria\\
         Emails: \email{sertl@hep.itp.tuwien.ac.at}, \email{grumil@hep.itp.tuwien.ac.at}, \email{niklasj@hep.itp.tuwien.ac.at}}

\abstract{

We correct a sign in the first variation of the on-shell action of cosmological topologically massive gravity at the chiral point and present the three equations affected by that sign. While this does not change any of the main conclusions of \href{http://arxiv.org/abs/0805.2610}{{\tt 0805.2610}}, it modifies the finite part of the Brown--York stress tensor. Our corrected Brown--York stress tensor is still finite, conserved and traceless, but no longer coincides with that of global AdS$_3$. It agrees with results found in recent literature.

}


\preprint{TUW-09-16}

\begin{document}

For simpler orientation we use the same notation and the same equation labels as in \cite{Grumiller:2008qz}. The equations below correct the corresponding ones in that work.

The first variation of the on-shell action is given by

\begin{multline}
\de I_{\rm CG}\big|_{\rm EOM} = -\int_{\dM}\!\!\!\!d^2x \sqrt{-\ga}\,\Big(K^{ij}-\big(K-\frac1\ell\big)\ga^{ij}\Big)\de\ga_{ij}\\
+\ell\int_{\dM}\!\!\!\!d^2x\,\epsilon^{ij}\,\Big(-R^{k\rho}{}_{j\rho}\,\de\ga_{ik}+K_i{}^k\,\de K_{kj}-\frac12 \Ga^k{}_{li}\,\de\Ga^l{}_{kj}\Big)
\tag{4.1}
\end{multline}

Note the relative minus sign in front of the first term in the second line. This sign change modifies also various other formulas, which we collect below. The last formula in the appendix must be corrected as follows:

\begin{multline}
\de I_{\rm CG}\big|_{\rm EOM} = \lim_{\rho\to\infty}\int_{\dM}\!\!\!\!d^2x\sqrt{-\ga^{(0)}}\,\de\ga_{ij}^{(0)}\, \Big[\frac{\rho}{\ell}\big(\ga^{ij}_{(1)}-\ga^{il}_{(1)}\ga_{lk}^{(0)}\eps^{kj}\big) -\frac{1}{2\ell}\,\big(\ga^{ij}_{(1)}-3\ga^{il}_{(1)}\ga_{lk}^{(0)}\eps^{kj}\big) \\
 +\frac1\ell\,\big(\ga_{(2)}^{ij}-\ga^{il}_{(2)}\ga_{lk}^{(0)}\eps^{kj}\big)
\Big]
\tag{A.16}
\end{multline}

The divergent piece in the first line and the finite part containing $\ga_{(2)}$ in the second line are unchanged, as well as the term proportional to $\ga^{ij}_{(1)}$. However, the term proportional to $\ga^{il}_{(1)}\ga_{lk}^{(0)}\eps^{kj}$ now has a different numerical factor. As an important consequence the $\ga_{(1)}$-terms no longer cancel on-shell.

This has the following implications for the results in section 4. The Brown--York stress tensor is now given by

\begin{equation}
T^{ij} = \lim_{\rho\to\infty} \frac1\ell \,\Big[\rho\,\big(\ga^{ij}_{(1)}-\ga^{il}_{(1)}\ga_{lk}^{(0)}\eps^{kj}\big) -\frac12\,\big(\ga^{ij}_{(1)}-3\ga^{il}_{(1)}\ga_{lk}^{(0)}\eps^{kj}\big) + \ga_{(2)}^{ij} - \ga^{il}_{(2)}\ga_{lk}^{(0)}\eps^{kj} \Big] +(i\leftrightarrow j)
\tag{4.8}
\end{equation}

By virtue of (4.5)-(4.7), (4.9) and reinserting $16\pi G$ it simplifies to

\begin{equation}
T^{ij} = -\frac{1}{\pi G\,\ell^3} \, \left(\begin{array}{cc}
1 & 1 \\
1 & 1 
\end{array}\right)^{\!ij} - \frac{2}{\pi G\,\ell^5} \,\cos{(2u)} \, \left(\begin{array}{rr}
1 & -1 \\
-1 & 1 
\end{array}\right)^{\!ij}
\tag{4.10}
\end{equation}

The result above implies that the Brown--York stress tensor is finite, conserved and traceless as in \cite{Grumiller:2008qz}. However, it does not coincide with the one from global AdS$_3$ as stated in \cite{Grumiller:2008qz}.\footnote{It is quite interesting that gravitational waves like the logarithmic modes backreact on the metric in such a way that they contribute in leading order to the Brown--York stress tensor $T_{ij}$. This is in contrast to what happens in general relativity.} 

In particular it is not chiral and depends on the light-like coordinate $u=\tau+\phi$.

Another relevant consequence is that the formulas (14) and (16) in \cite{Grumiller:2008es} are modified. Again this does not alter the main conclusions of that work, which establishes consistent boundary conditions for CTMG at the chiral point, but it leads to non-chirality of the generators of the asymptotic symmetry group.

Based upon the constraint analysis by Steve Carlip \cite{Carlip:2008qh} (which agrees with the earlier analysis performed at the chiral point \cite{Grumiller:2008pr}) the correct result for the asymptotic charges was obtained first by Marc Henneaux, Christian Martinez and Ricardo Troncoso \cite{Henneaux:2009pw}, and shortly afterwards by Alex Maloney, Wei Song and Andy Strominger \cite{Maloney:2009ck}. The correct result for the Brown--York stress tensor was obtained first by Kostas Skenderis, Marika Taylor and Balt van Rees \cite{Skenderis:2009nt}. Up to notational issues our results above agree with the results of these authors.

\acknowledgments

We thank the participants of the workshop `Gravity in three dimensions' at the Erwin-Schr\"odinger Institute (ESI) for discussions.

SE, DG and NJ are supported by the START project Y435-N16 of the Austrian Science Foundation (FWF). NJ acknowledges financial support from ESI.

%




\begin{thebibliography}{1}



\bibitem{Grumiller:2008qz}
D.~Grumiller and N.~Johansson, {\it {Instability in cosmological topologically
  massive gravity at the chiral point}},  {\em JHEP} {\bf 07} (2008) 134
  [\href{http://arXiv.org/abs/0805.2610}{{\tt 0805.2610v2}}].



\bibitem{Grumiller:2008es}
D.~Grumiller and N.~Johansson, {\it {Consistent boundary conditions for
  cosmological topologically massive gravity at the chiral point}},  {\em Int.
  J. Mod. Phys.} {\bf D17} (2009) 2367--2372
  [\href{http://arXiv.org/abs/0808.2575}{{\tt 0808.2575v3}}].



\bibitem{Carlip:2008qh}
S.~Carlip, {\it {The Constraint Algebra of Topologically Massive AdS Gravity}},
   {\em JHEP} {\bf 10} (2008) 078 [\href{http://arXiv.org/abs/0807.4152}{{\tt
  0807.4152}}].




\bibitem{Grumiller:2008pr}
D.~Grumiller, R.~Jackiw and N.~Johansson, {\it {Canonical analysis of
  cosmological topologically massive gravity at the chiral point}}, ch.~20 in ``Fundamental Interactions -- A Memorial Volume for Wolfgang Kummer'', World Scientific (2009) [\href{http://arXiv.org/abs/0806.4185}{{\tt 0806.4185}}].




\bibitem{Henneaux:2009pw}
M.~Henneaux, C.~Martinez and R.~Troncoso, {\it {Asymptotically anti-de Sitter
  spacetimes in topologically massive gravity}},  {\em Phys. Rev.} {\bf D79}
  (2009) 081502R [\href{http://arXiv.org/abs/0901.2874}{{\tt 0901.2874}}].




\bibitem{Maloney:2009ck}
A.~Maloney, W.~Song and A.~Strominger, {\it {Chiral Gravity, Log Gravity and
  Extremal CFT}},  \href{http://arXiv.org/abs/0903.4573}{{\tt 0903.4573}}.




\bibitem{Skenderis:2009nt}
K.~Skenderis, M.~Taylor and B.~C. van Rees, {\it {Topologically Massive Gravity
  and the AdS/CFT Correspondence}},  \href{http://arXiv.org/abs/0906.4926}{{\tt
  0906.4926}}.




\end{thebibliography}

\providecommand{\href}[2]{#2}\begingroup\raggedright\endgroup

\end{document}